\documentclass[journal,twoside,web]{ieeecolor}
\usepackage{tmi}
\usepackage{cite}
\usepackage{amsmath,amssymb,amsfonts}
\usepackage{graphicx}
\usepackage{textcomp}

\usepackage{todonotes}
\usepackage[colorlinks=true]{hyperref}
\usepackage{mathtools}
\usepackage{amsmath}
\usepackage{amsfonts}

\usepackage{latexsym,graphicx,array,amsfonts}

\usepackage{multirow}
\usepackage{algorithm}
\usepackage[noend]{algpseudocode}
\usepackage{array}
\newcolumntype{L}[1]{>{\raggedright\let\newline\\\arraybackslash\hspace{0pt}}m{#1}}
\newcolumntype{C}[1]{>{\centering\let\newline\\\arraybackslash\hspace{0pt}}m{#1}}
\newcolumntype{R}[1]{>{\raggedleft\let\newline\\\arraybackslash\hspace{0pt}}m{#1}}
\usepackage[font=small,skip=5pt]{caption}

\usepackage{booktabs} 
\usepackage{colortbl}

\newcommand{\mc}[3]{\multicolumn{#1}{#2}{#3}}
\definecolor{Gray}{gray}{0.85}
\definecolor{LightCyan}{rgb}{0.88,1,1}

\newcolumntype{a}{>{\columncolor{Gray}}c}
\newcolumntype{b}{>{\columncolor{white}}c}

\newcommand{\inline}{\mbox}

\begin{document}
\title{Label Refinement Network from Synthetic Error Augmentation for Medical Image Segmentation}

\author{Shuai~Chen,
        Antonio~Garcia-Uceda,
        Jiahang~Su,
        Gijs van Tulder,
        Lennard Wolff,
        Theo~van~Walsum,
        ~Marleen~de~Bruijne,~\IEEEmembership{Member,~IEEE,}
        and on behalf of the MR CLEAN Registry investigators
\thanks{Manuscript received May 31, 2022. This work was partially funded by Chinese Scholarship Council (File No.201706170040), Netherlands Organisation for Scientific Research (NWO) project VI.C.182.042, the MARBLE project (EFRO/OP-Oost: PROJ-00887), the Contrast project (Dutch Heart Foundation (CVON2015-01: CONTRAST), the Brain Foundation Netherlands (HA2015.01.06) and additional funding by the Ministry of Economic Affairs by means of the PPP Allowance made available by the Top Sector Life Sciences \& Health to stimulate public-private partnerships (LSHM17016)).}%
\thanks{This work involved medical images from human subjects acquired for clinical studies. Approval of the study protocols was granted by the central medical ethics committee of the hospital Erasmus MC, Rotterdam, the Netherlands (No. MEC-2013‐338 and MEC-2014-235).}%
\thanks{S. Chen, A. Garcia-Uceda, and J. Su have equal contributions.}%
\thanks{S. Chen, A. Garcia-Uceda, J. Su, T. van Walsum, and M. de Bruijne are with the Biomedical Imaging Group Rotterdam, Department of Radiology~\&~Nuclear Medicine, Erasmus MC, Rotterdam, The Netherlands (e-mail: s.chen.2@erasmusmc.nl, a.garciauceda@erasmusmc.nl, j.su@erasmusmc.nl, t.vanwalsum@erasmusmc.nl, marleen.debruijne@erasmusmc.nl).}%
\thanks{G. van Tulder is with the Data Science group, Faculty of Science, Radboud University, Nijmegen, The Netherlands (e-mail: g.vantulder@cs.ru.nl).}%
\thanks{L. Wolff is with Department of Radiology \& Nuclear Medicine, Erasmus MC, Rotterdam, The Netherlands (e-mail: l.wolff.1@erasmusmc.nl).}%
\thanks{M. de Bruijne is with the Machine Learning Section, Department of Computer Science, University of Copenhagen, DK-2110 Copenhagen, Denmark.}}

% The paper headers
\def\BibTeX{{\rm B\kern-.05em{\sc i\kern-.025em b}\kern-.08em
    T\kern-.1667em\lower.7ex\hbox{E}\kern-.125emX}}
\markboth{\journalname, VOL. XX, NO. XX, XXXX 2022}
{Chen \MakeLowercase{\textit{et al.}}: Label Refinement Network from Synthetic Error Augmentation for Medical Image Segmentation}

% make the title area
\maketitle

\begin{abstract}
Deep convolutional neural networks for image segmentation do not learn the label structure explicitly and may produce segmentations with an incorrect structure, e.g., with disconnected cylindrical structures in the segmentation of tree-like structures such as airways or blood vessels. In this paper, we propose a novel label refinement method to correct such errors from an initial segmentation, implicitly incorporating information about label structure. This method features two novel parts: 1) a model that generates synthetic structural errors, and 2) a label appearance simulation network that produces synthetic segmentations (with errors) that are similar in appearance to the real initial segmentations. Using these synthetic segmentations and the original images, the label refinement network is trained to correct errors and improve the initial segmentations. The proposed method is validated on two segmentation tasks: airway segmentation from chest computed tomography (CT) scans and brain vessel segmentation from 3D CT angiography (CTA) images of the brain. In both applications, our method significantly outperformed a standard 3D U-Net and other previous refinement approaches. Improvements are even larger when additional unlabeled data is used for model training. In an ablation study, we demonstrate the value of the different components of the proposed method.

\end{abstract}


\section{Introduction}
\label{sec:introduction}

\begin{figure}[t]
    \centering
    % IMPORTANT: the .pdf image failed when generating the image in the PDF proof of TMI. I converted to .png to solve this during submission. For the final publication we can use .pdf
    \includegraphics[width=\linewidth]{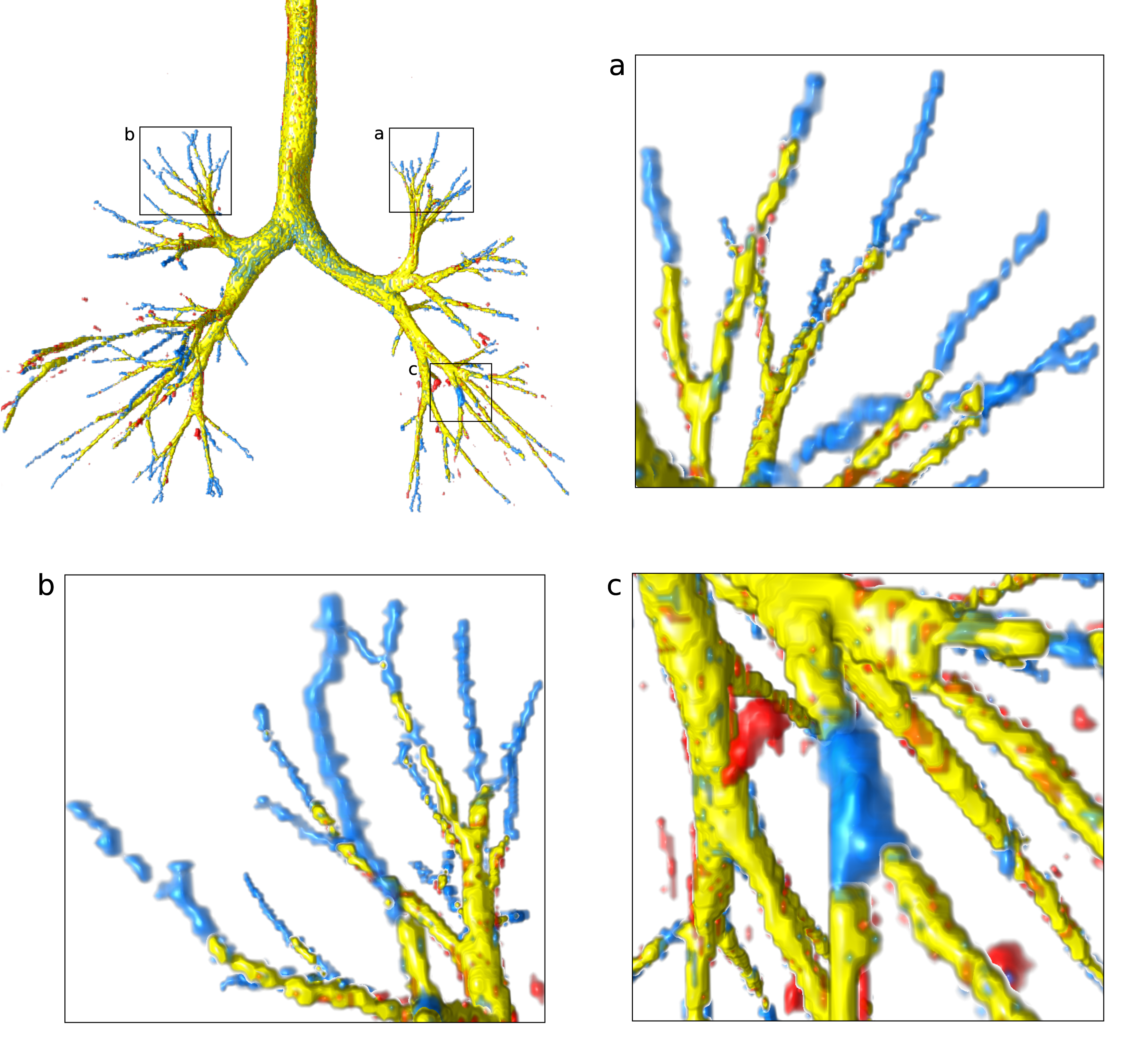}
    \caption{\textbf{Common structural errors in the segmentations obtained by a U-Net, trained to segment airways in the lungs~\cite{garciauceda2021}.} True positives are displayed in yellow, false negatives in blue and false positives in red. Detailed views a-b show errors as missing terminal branches, and view c shows a discontinuity error in the branch. Better to view in zoom in color.}
    \label{fig:problems}
\end{figure}

Convolutional neural networks (CNNs) are the state-of-the-art for many biomedical imaging segmentation tasks. Many CNN segmentation architectures have been proposed, such as fully connected networks~\cite{long2015fully}, Dense-Net~\cite{huang2017densely} and the U-Net~\cite{ronneberger2015u}. The U-Net has become the most popular network for biomedical image segmentation, due to its efficient structural design featuring skip-connections, showing superior accuracy and robustness in various segmentation tasks~\cite{UNetreview,isensee2021nnu}. Most CNN-based segmentation methods including the U-Net do not fully exploit and encode the structural information of the objects to be segmented. Consequently, these methods may produce segmentations with errors that become obvious when looking at the full segmented structure. Examples of such errors are discontinuities in the segmentations of elongated tubular structures, such as airways in the lungs, as shown in Figure~\ref{fig:problems}. Using label structural knowledge such as continuity in the branches of the airway tree can help prevent these errors. However, it is not trivial to explicitly encode this global information in CNNs.

In this paper, we propose a framework to implicitly encode the label structural information into CNNs by formulating this as a label refinement step. Specifically, we generate structural errors in labels (such as the ground truth or initial segmentations) and train a label refinement network to correct these errors. The trained network is expected to generalize to the real errors in the initial segmentations produced by a baseline segmentation network and correct them. To enhance the generalizability of the label refinement network on the initial segmentations, a label appearance simulation network is applied to reduce the appearance difference between the synthetic labels and the initial segmentations. With these synthetic labels (and the initial segmentations) together with the original image as inputs and the ground truth segmentations as reference, the label refinement network can learn to correct those errors and incorporate this in its segmentation decisions.

We validated the proposed label refinement method on two segmentation tasks: airway segmentation from chest computed tomography (CT) scans~\cite{garciauceda2021} and brain vessel segmentation from 3D CT angiography (CTA) images of the brain~\cite{jsu}. We compared our method with a U-Net baseline and other refinement networks, including DoubleU-Net~\cite{doubleunet} and SCAN~\cite{adv}, an adversarial refinement network. Moreover, we conducted an ablation study to show the contribution of each individual component of the label refinement method. Finally, we performed experiments in a semi-supervised setting to train our method using additional unlabeled data.

\begin{figure*}[t]
    \centering
    \includegraphics[width=\textwidth]{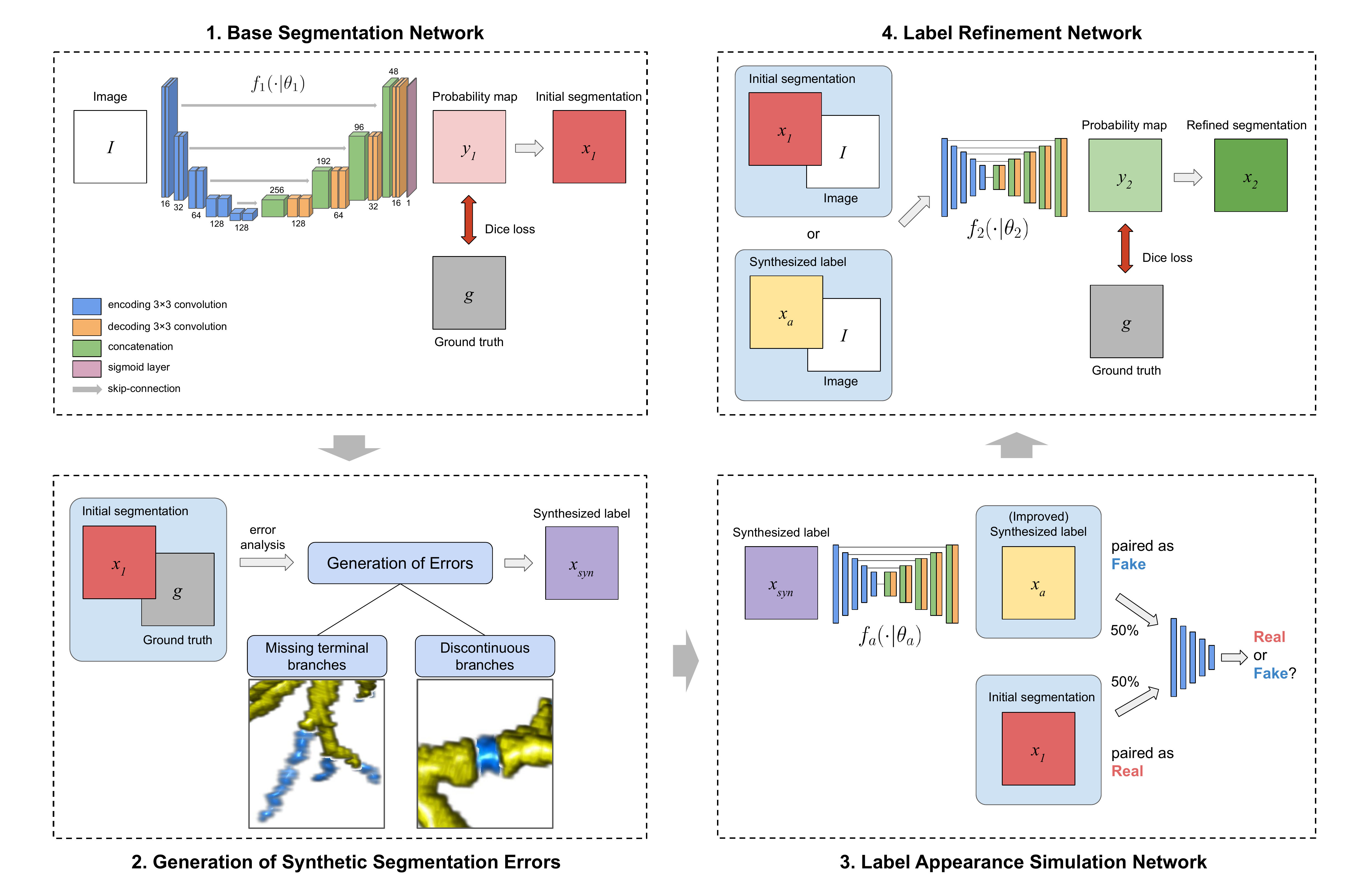}
    \caption{\textbf{Schematics of the proposed label refinement method.} First, a base segmentation network $f_{1}$ is trained to obtain the initial predictions $y_{1}$. Second, we synthesize a new dataset ${y}_{\mathrm{n}}$ that contains similar errors to $y_{1}$. Third, a label appearance improvement network $f_{\mathrm{a}}$ (together with a discriminator $D$) is trained to obtain a more realistic dataset $y_{\mathrm{a}}$. Finally, the label refinement network $f_{2}$ is trained with $y_{1}$ and $y_{\mathrm{a}}$ together with the image $I$ as inputs.}
    \label{fig:main}
\end{figure*}

\section{Related Work}
\label{sec:relatedwork}

\subsection{Label Refinement}
\label{RWlabelrefinement}

In this work, we apply a refinement network on the initial segmentation from a baseline segmentation network together with the original image, with the aim of correcting errors in the initial segmentation. A similar approach has been used in other previous papers. Jha~et~al.~\cite{doubleunet} attached a second U-Net network to a baseline U-Net, using as inputs the original image multiplied with the output of the first U-Net. Yang~et~al.~\cite{labelrefine} refined low-quality manual annotations made by non-experts by training their method with added noise in order to reduce the inter-observer inconsistency of the annotations. Unlike our method, Yang~et~al. do not focus on refining an initial automatic segmentation and therefore the label appearance simulation network is not needed. Dai~et~al.~\cite{adv} refined the segmentations from a fully convolutional network by using adversarial training to reduce the domain gap between the target predictions and the ground truth segmentations on training data. Ara\'{u}jo~et~al.~\cite{araujo} attached a variational auto-encoder after a U-Net network to encode the label topology of the ground truth segmentations for a better label reconstruction. Different from Dai~et~al. and Ara\'{u}jo~et~al., our work does not focus on encoding~\cite{araujo} or discriminating~\cite{adv} the overall label topology, but instead on learning to correct the most common errors in the segmentations.

\subsection{Airway Segmentation}
\label{RWairway}

The airway tree in the lungs forms a complex 3D tree-like branching network, with many branches of different sizes and orientations. The peripheral branches of smaller size are challenging to segment from chest CT scans, as they have obscured borders due to partial volume effects. Many classical methods for airway tree extraction are based on a region growing algorithm~\cite{graham2010,lo2010,lo2009}. However, their accuracy is limited, and they typically miss a large number of the smaller peripheral airways~\cite{exact2012}. Many state-of-the-art airway segmentation methods are based on CNNs, and especially the U-Net~\cite{garciauceda2021,qin2021,cheng2021,zheng2021}. CNN-based methods can obtain more accurate and complete segmentations than previous intensity-based methods. However, even the latest U-Net-based methods usually miss several terminal branches, and make errors in continuity around the smaller segmented branches.

\subsection{Brain Vessel Segmentation} 
\label{RWvessel}

The brain vessels form a complex 3D branching network that consists of veins and arteries. In 3D CTA images of the brain, many seemingly isolated vessel structures can be present due to the image acquisition and vascular diseases, such as ischemic large vessel occlusions. State-of-the-art vessel segmentation methods have been applied to 3D time-of-flight (TOF) magnetic resonance angiography (MRA) images~\cite{sanches,livne,adam}, and to 3D and 4D CTA images~\cite{meijs} using U-Nets. Su~et~al.~\cite{jsu} used a U-Net-based method to extract a dilated vessel centerline approximation. Compared to previous vessel segmentation methods~\cite{sanches,livne,adam, meijs}, centerline extraction recovers the topology of the vessel structure more accurately (e.g., ``kissing vessels'' appear connected in the full segmentations but are disconnected in centerline extraction). However, the U-Net still makes other topological errors such as local connectivity gaps in vessel branches.

\section{Method}

\subsection{Overview}
\label{sec:overview}

The proposed method consists of four steps, schematically shown in Figure~\ref{fig:main}. Firstly, a baseline segmentation network generates the initial segmentations (Section~\ref{sec:segmentation}). Secondly, synthetic errors are generated and added to every ground truth segmentation, in order to generate synthetic labels to train the label refinement network (Section~\ref{sec:errors}). Thirdly, a label appearance simulation network (LASN) based on adversarial learning is used to reduce the appearance difference between the synthetic labels and the initial segmentations (Section~\ref{sec:gan}). Finally, a label refinement network is trained to predict the final segmentation, using the synthetic labels (or the initial segmentations) and the original images as inputs, and the ground truth segmentations as reference (Section~\ref{sec:refinement}).

\subsection{Base Segmentation Network}
\label{sec:segmentation}

We use a base segmentation network $f_{1}$ to predict an initial segmentation. Given a medical imaging dataset that contains an image $I$ and the ground truth segmentation $g$ for each subject, the model $f_{1}(I|\theta_{1})$, with $\theta_{1}$ the trainable parameters, is trained by minimizing the Dice loss $\mathcal{L}_{1}=\mathcal{L}_{\mathrm{dc}}(f_{1}(I),g)$:

\begin{equation}
\mathcal{L}_{\mathrm{dc}}(y,g) = -\frac{{2}\sum_{i\in{I}}y_{i}g_{i}}{\sum_{i\in{I}}y_{i}+\sum_{i\in{I}}g_{i}}
\label{eq:diceloss}
\end{equation}
where $y_{i}$ and $g_{i}$ are the $i$th voxel values of the probability maps output by the model (in this case $y=f_{1}(I)$), and the ground truth segmentation, respectively.

The initial predicted segmentation is $x_{1}$, obtained by thresholding the output probability maps $y_{1}$ of the network with value 0.5. $x_{1}$ may contain label structural errors, such as discontinuous branches in a tree-like structure. Next, we show how to design synthetic errors similar to those in $x_{1}$ that can be used to train the label refinement network.

\subsection{Generation of Synthetic Segmentation Errors}
\label{sec:errors}

We use synthetic labels $\mathbf{x}_{\mathrm{syn}}$ with added synthetic errors to train the label refinement network. Depending on the experimental set-up, the errors can be added to the ground truth or to the initial segmentations. The synthetic errors are generated to resemble those in the initial segmentations $x_{1}$, based on our initial analysis of common errors. In this paper, we focus on two structures: airways in the lungs and vessels in the brain. Airways and vessels share several characteristics: they both form 3D branching networks, with branches of cylindrical shape and various sizes and orientations. We use this prior shape knowledge to generate synthetic errors, as described below for each structure.

\subsubsection{Synthetic Errors For Airways}
\label{sec:airwayerror}

Most of the errors in airway segmentations can be grouped into two types: 1) missing terminal branches, partially or totally, and 2) discontinuity in the segmented branches, which occurs more frequently in smaller branches. Examples of errors in airway segmentations obtained by the baseline segmentation network in Section~\ref{sec:segmentation} are shown in Figure~\ref{fig:problems}. To generate similar, synthetic errors, we select a random subset of branches in the airway tree and partially remove the segmentation of the selected branches by masking it at a random position and with a random length. Branches are identified using the airway centerline tree, extracted from the airway segmentation~\cite{exact2012}. Single branches are defined as the segments between two bifurcation points or between the last bifurcation and the end of terminal branches. The applied masking is defined differently for each type of error:  

\emph{Missing terminal branches}: The subset of branches in which to synthesize errors is randomly sampled from all the terminal branches in the airway tree, defined as branches with no further bifurcations downstream. A mask of cylindrical shape is applied to (partially) remove the selected branch. The mask is defined by 1) a starting point, that is a random position along the branch centerline between the branch start and middle points; 2) a length, that is the distance between the mask start point and branch end; and 3) a width, that is three times the branch diameter.

\emph{Discontinuity in branches}: The subset of branches with errors is randomly sampled from all the branches in the airway tree, excluding the trachea, the two main bronchi and the 2\textsuperscript{nd} generation airways, and including the terminal branches. We assign a higher sampling probability to branches of higher airway generation, where the generation is defined as the number of branch bifurcations counted in the path linking the given branch and the root of the airway tree, i.e., the trachea. The sampling probability $p_{i}$ for each candidate branch is defined as $p_{i}=g_{i}/\sum\nolimits_{k=1}^{N_c}{g_{k}}, \forall i=1\dots N_{c}$, where $g_{i}$ is the airway generation and $N_{c}$ the number of candidate branches. A mask of cylindrical shape is applied to create a gap in the selected branch. The mask is defined by 1) a center, that is a random position along the branch centerline; 2) a length, that is a random distance between a minimum of 10 voxels and the total branch length; and 3) a width, that is three times the branch diameter.

\emph{Parameters}: The extent of each type of errors in the airway synthetic labels is determined by a separate parameter, denoted as $p^{\mathrm{a}}_1$ and $p^{\mathrm{a}}_2$. $p^{\mathrm{a}}_1$ is the proportion of selected branches with errors of type ``missing terminal branches'', with respect to all the terminal branches. $p^{\mathrm{a}}_2$ is the proportion of selected branches with errors of type ``discontinuity in branches'', with respect to all the branches in the airway tree.

\subsubsection{Synthetic Errors for Brain Vessels}
\label{sec:vesselerror}

Most of the errors in brain vessel segmentation are in the form of incomplete or missing vessel branches. To generate similar, synthetic errors, we create random discontinuous gaps in the segmentation of each vessel by masking it at a random position and with a random length. Since the errors occur more frequently for long vessels than for short ones, we group all the vessels into three equal-sized groups: long, medium size and short, based on the relative centerline segment lengths in each subject. The distribution of vessel lengths (in voxels), using the median and interquartile range (IQR), is: for long segments \inline{70~(49--106)}, for medium-size segments \inline{29~(22--36)} and for short segments \inline{13~(9--17)}. For long segments, the maximum number of injected gaps is 6 (randomly sampled from a uniform distribution between 0 and 6 positions) with gap length between \inline{10--35} voxels. For medium-size segments, the maximum number of gaps is 4 with gap length between \inline{10--20} voxels. For the short segments, the maximum number of gaps is 2 with gap length between \inline{6--15} voxels. Those error injections are applied on the 1 voxel-wide ground truth centerlines, by dilating it with a 3$\times$3$\times$3 cubic structure element to generate the final vessel synthetic label.

\emph{Parameters}: The extent of errors in the vessel synthetic labels is determined by only one parameter, denoted as $p^{\mathrm{v}}$. $p^{\mathrm{v}}$ is the proportion of selected branches with errors with respect to all the branches in the vessel network.

\subsection{Label Appearance Simulation Network}
\label{sec:gan}

\begin{figure}[t]
    \centering
    \includegraphics[width=\linewidth]{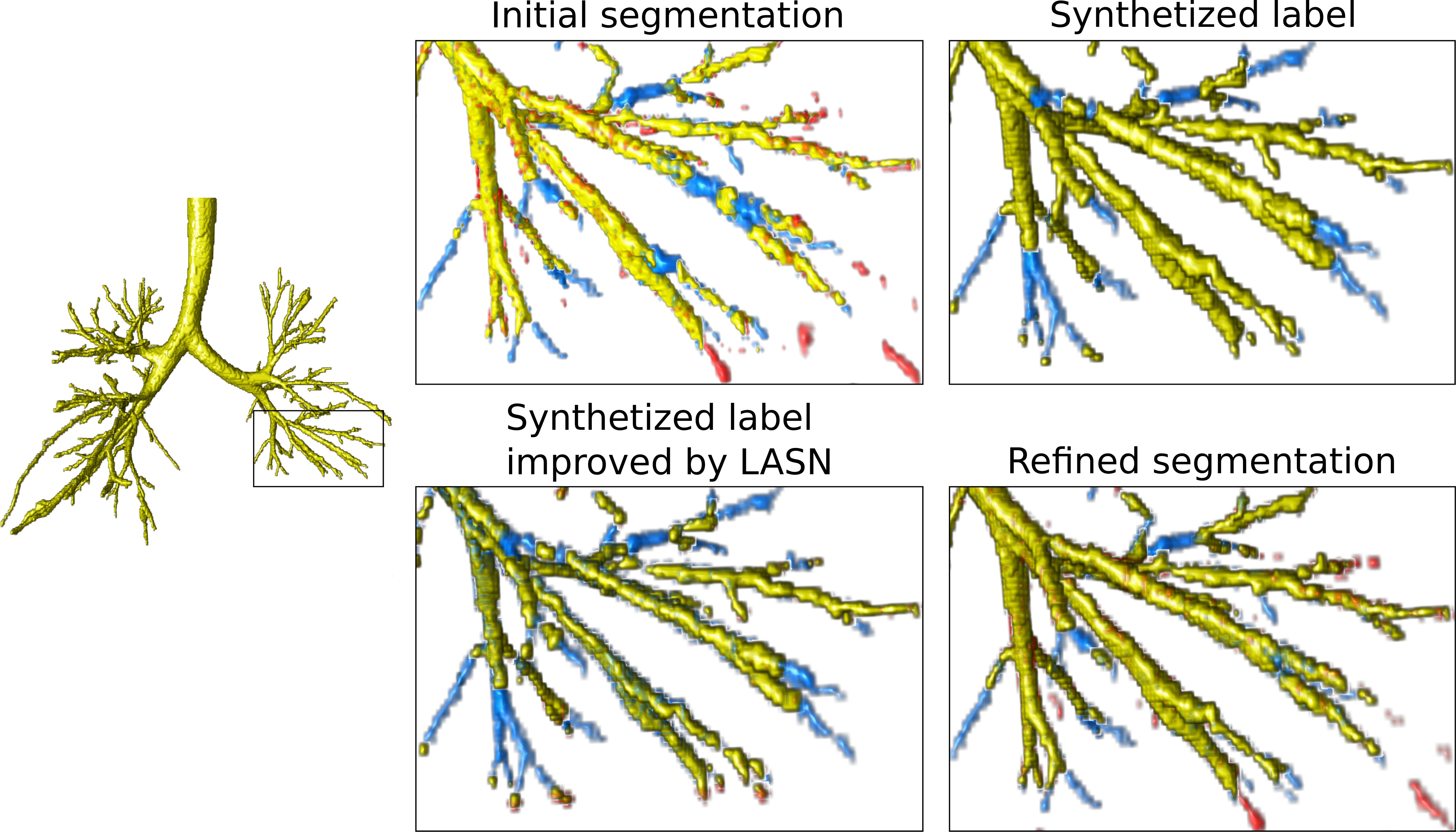}
    \caption{\textbf{Example of segmentation of airways in the lungs obtained by the different components of the proposed method.} In the detailed views, true positives are displayed in yellow, false negatives in blue and false positives in red. Better to view in zoom in color.}
    \label{fig:example}
\end{figure}

Although the synthetic labels $x_{\mathrm{syn}}$ are designed to have similar structural errors to the initial segmentation $x_{1}$, there may be an appearance difference between $x_{1}$ and $x_{\mathrm{syn}}$ (see an example in Figure~\ref{fig:example}a and b). The label refinement network trained on $x_{\mathrm{syn}}$ may therefore generalize poorly to $x_{1}$. To prevent this, we use a label appearance simulation network $f_{\mathrm{a}}(\cdot|\theta_{\mathrm{a}})$ to change the appearance of $x_{\mathrm{syn}}$ to be more similar to that of $x_{1}$, while preserving the synthetic errors that we added to $x_{\mathrm{syn}}$. 

The label appearance simulation network $f_{\mathrm{a}}(\cdot|\theta_{\mathrm{a}})$, with $\theta_{\mathrm{a}}$ the trainable parameters, is optimized by adversarial learning via a discriminator $D$:

\begin{equation}
\label{equ:appearencenetwork}
f_{\mathrm{a}}^{*}=\arg\min\limits_{f_{\mathrm{a}}}((\max\limits_{D}\mathcal{L}_{\mathrm{adv}}(f_{\mathrm{a}},D)) + \lambda\mathcal{L}_{\mathrm{dc}}(x_{\mathrm{a}},x_{\mathrm{syn}}))
\end{equation}
with the adversarial loss $\mathcal{L}_{\mathrm{adv}}$ defined as:

\begin{equation}
\mathcal{L}_{\mathrm{adv}}(f_{\mathrm{a}},D)=\mathbb{E}_{x_{1}}[\text{log}{D}(x_{1})] + \mathbb{E}_{x_{\mathrm{a}}}[\text{log}(1-{D}(x_{\mathrm{a}}))] 
\end{equation}
where $D$ is a classifier, discriminating the given label $x$ and the initial segmentation $x_{1}$. It outputs a probability between 0.0 and 1.0. $x_{\mathrm{a}}=f_{\mathrm{a}}(x_{\mathrm{syn}})$ is the appearance-enhanced label of $x_{\mathrm{syn}}$. We added a Dice-based identity loss $\mathcal{L}_{\mathrm{dc}}(x_{\mathrm{a}}, x_{\mathrm{syn}})$ to train $f_{\mathrm{a}}(\cdot)$, in order to preserve the synthetic errors that we added in $x_{\mathrm{syn}}$. The hyperparameter $\lambda$ controls the balance between the identity loss and the dissimilarity adversarial loss.

\subsection{Label Refinement Network}
\label{sec:refinement}
Finally, we optimize a label refinement network $f_{2}$ to predict the ground truth segmentations, based on the synthetic labels with errors ${x}_{\mathrm{a}}$ together with the original image as inputs. This way, $f_{2}$ learns to correct segmentation errors and can be used to improve the initial segmentations $x_{1}$. The model $f_{2}(I|\theta_{2})$, with $\theta_{2}$ the trainable parameters, is trained by minimizing the Dice loss $\mathcal{L}_{2}=\mathcal{L}_{\mathrm{dc}}(f_{2}(I,\Tilde{x}),g)$, given by equation~\eqref{eq:diceloss}. The final segmentation result is $x_{2}$, obtained by thresholding the output probability maps $y_{2}$ of the refinement network with value 0.5.

\section{Experiments}
\label{sec:experiments}

\subsection{Datasets}
\label{Data}

We validated the proposed method on two biomedical imaging segmentation tasks: segmenting airways from chest CT scans and brain vessels from CTA images of the brain.

\subsubsection{Chest CT data}
\label{DataAirways}

The dataset of chest CT scans is from a retrospective study of pediatric patients (6 to 17 years old) with cystic fibrosis lung disease, acquired routinely at the hospital Erasmus MC-Sophia Rotterdam~\cite{chestCTdata}. The CT scans show noticeable structural airway abnormalities resulting from the disease. In our study, we used 178 low-dose CT scans acquired at full inspiration breath-hold. All CT scans have slice dimensions 512$\times$512, with a variable number of slices between \inline{200--1000}. Each CT scan has an in-plane voxel size in the range \inline{0.35--0.65}~mm, with slice thickness between \inline{0.75--1.0}~mm, and slice spacing between \inline{0.3--0.8}~mm. A random subset of 65 CT scans from the total 178 scans have annotations of the airway lumen. To obtain these annotations, Thirona's lung quantification software LungQ (Thirona, Nijmegen, the Netherlands) was used to automatically extract the airway lumen from the CT scan. Then, these segmentations were visually checked by trained data analysts for accuracy, and corrected as needed.

For our experiments, we used as testing data 41 random CT scans from the subset of 65 CT scans with ground truth segmentations. From the remaining 24 CT scans with annotations, we used three different random data splits with 20 CT scans for training the networks and 4 CT scans for validation. The remaining 113 CT scans without ground truth segmentations were used as unlabeled training data for the experiments with semi-supervised learning.

\subsubsection{CTA data of the Brain}
\label{DataVessels}

The dataset of CTA images of the brain is from the MR CLEAN Registry~\cite{jansen}, an ongoing registry for patients who underwent endovascular treatment for acute ischemic stroke in one of 19 hospitals in the Netherlands since March 2014. The data was collected during clinical practice, and we applied the following data inclusion criteria: 1) slice thickness $\leq$1.5~mm, 2) slice spacing $\leq$1.5~mm, 3) the contrast acquisition phase has to be peak arterial phase, equilibrium or early venous phase~\cite{rodriguez2014venous}, and 4) the image should cover at least half of the brain. In our study, we used 69 CTA images from 69 different subjects used in~\cite{jsu}. All CTA images were skull-stripped with an atlas-based registration method~\cite{roman}. 20 CTA images had no vessel annotations, 9 CTA images had a complete brain vessel centerline annotation, and the remaining 40 CTA images (randomly sampled from the whole dataset) had vessel centerline annotations in a randomly sampled sub-volume of 140$\times$140$\times$140 voxels. The centerline annotations were dilated with a 3$\times$3$\times$3 cubic structure element to obtain the ground truth segmentations. Each CTA image has an in-plane voxel size in the range \inline{0.4--0.68}~mm, with slice thickness between \inline{0.5--1.5}~mm, and slice spacing between \inline{0.3--1.0}~mm.

For our experiments, we used as testing data 2 random full-volume CTA scans and 20 random CTA cubes from the set of 9 full-volume, annotated CTA scans and 40 CTA cubes, respectively. From the remaining data with annotations, we used three different random data splits with 7 full-volume CTA scans and 14 CTA cubes for training the networks, and 6 CTA cubes for validation. The remaining 20 full-volume CTA scans without manual annotations were used as unlabeled training data for the experiments with semi-supervised learning.

\subsection{Parameters for Generating Synthetic Errors}
\label{sec:parameters}

The generation of synthetic errors depends on the parameters $p^{\mathrm{a}}_1$ and $p^{\mathrm{a}}_2$ for airways, and $p^{\mathrm{v}}$ for vessels, described in Sections~\ref{sec:airwayerror} and~\ref{sec:vesselerror} respectively. In the rest of the paper we refer to these parameters as ``synthetic error rate'', for each type of error. For each training sample, the synthetic error rate is randomly sampled from a uniform distribution between 0.0 and the upper bound, or maximum synthetic error rate. These upper bounds are hyperparameters for the proposed method, denoted as $P^{\mathrm{a}}_1$ and $P^{\mathrm{a}}_2$ for airways, and $P^{\mathrm{v}}$ for vessels.

We conducted experiments varying the hyperparameters for the error generation in the proposed method, i.e., the maximum synthetic error rates ($P^{\mathrm{a}}_1$ and $P^{\mathrm{a}}_2$ for airways, and $P^{\mathrm{v}}$ for vessels), to investigate their influence in the method performance. The results are shown in Section~\ref{sec:influence} below.

In our further experiments, the optimal hyperparameters were determined on the validation set for each of the three random data splits that we used, for both applications. Each hyperparameter was searched independently, from 0.0 to 1.0, while fixing the parameters for other error types to 0.0.

\subsection{Network Architecture}
\label{sec:ImplemNetworks}

The baseline segmentation network $f_{1}$ is a 3D U-Net~\cite{3dunet}, shown in Figure~\ref{fig:main}. The label refinement network $f_{2}$ and the label appearance simulation network $f_{\mathrm{a}}$ use a similar U-Net layout, with the discriminator $D$ in $f_{\mathrm{a}}$ using the same layout as the U-Net encoder. The U-Net consists of an encoding path followed by a decoding path, with skip-connections linking the two paths. The network has 5 levels of depth, 16 feature channels in the first layer, and an input image size of 128$\times$128$\times$128. Each level of the encoding / decoding paths consists of two 3$\times$3$\times$3 convolutional layers followed by a 2$\times$2$\times$2 pooling or upsampling layer, respectively. Each convolutional layer consists of 3$\times$3$\times$3 convolution with zero padding followed by instance normalization and leaky ReLU activation. The number of feature channels is doubled or halved after every pooling or upsampling layer, respectively. The last layer of the U-Net is a 1$\times$1$\times$1 convolution, combining the outputs into a single feature map, followed by a sigmoid activation. A training batch contains only one image due to GPU memory limits. The networks are implemented using PyTorch~\cite{pytorch}. The source code is publicly available: \url{https://github.com/ShuaiChenBIGR/Label-refinement-network}.

\subsection{Details of Training and Inference of Networks}
\label{sec:DetailTrainTest}

For training, we first apply random rigid transformations as data augmentation, in the form of 1) random 3D rotations up to 30 degrees for all axes, 2) random scaling with factor between \inline{0.7--1.4} and 3) random flipping in the three directions. Then, we generate samples by extracting random image patches of size 128$\times$128$\times$128 on the fly from the input training images and corresponding ground truth segmentations. For the airway segmentation experiments, a lung mask is applied to the output of the network and the ground truth patches before computing the training loss. For this operation, we use a pre-computed lung mask that is easily obtained with a region growing algorithm~\cite{lo2010}. During training, we used the Adam optimizer~\cite{adamoptime} with an initial learning rate of 1$\times10^{-2}$. To train the refinement network $f_{2}$, the label $\Tilde{x}$ in each training sample is randomly sampled with equal probability from either the initial segmentation $x_{1}$ or the synthetic label $x_{\mathrm{a}}$ after the label appearance simulation network.

During inference on new images, the input patches are extracted in a sliding-window fashion, with an overlap of 50\% in the three directions. Then, the patch-wise predicted output by the network is aggregated by stitching the patches together, to reconstruct the full-size segmentation result. For the airway segmentation experiments, we applied a lung mask to the final segmentation to remove any spurious noise prediction outside the lungs. For this operation, we use the same region growing algorithm as during training.

For the adversarial loss in equation~\eqref{equ:appearencenetwork}, the weight $\lambda$ is set to 0.01 for all experiments in this paper, based on visual inspection of the generated synthetic labels $x_{\mathrm{a}}$.

\subsection{Comparisons}

We compared the results of our proposed method with the baseline 3D U-Net segmentation network described in Section~\ref{sec:ImplemNetworks} (U-Net baseline). Additionally, we compared our method with two previous refinement approaches: DoubleU-Net~\cite{doubleunet} and SCAN~\cite{adv}. For both baselines, we reimplemented the methods from the original papers. The DoubleU-Net method consists of two consecutive U-Nets, with skip connections from the encoder of the first U-Net to the decoders of both U-Nets. The SCAN method uses a U-Net with a discriminator and adversarial loss, discriminating between the segmentation results and the ground truth. The weight for balancing the segmentation loss and the adversarial loss (low value on the adversarial term) is tuned between 0.001 and 0.1, on the validation sets for each application. For DoubleU-Net, no additional hyperparameters need to be tuned. Our implementations of DoubleU-Net and SCAN use the same 3D U-Net backbone as our proposed method and the first baseline.

We also conducted an ablation study of the proposed method (LR+Syn+LASN) by removing some of the components. We evaluated 1) a simple label refinement method by inputting the original images and the initial segmentations without any synthetic errors (LR), 2) a label refinement method with synthetic errors added to the initial segmentations (LR+Syn(init)), and 3) a label refinement method with synthetic errors added to the ground truth segmentations but without the label appearance simulation network (LR+Syn).

\subsection{Evaluation Metrics}

We evaluated the methods with the Dice coefficient to measure the overall segmentation quality, as well as with three metrics designed for tree-like structures: centerline completeness, centerline leakage, and number of gaps. For the airway segmentation experiments, the required centerlines were obtained by applying a skeletonization method~\cite{skeletonize1994} to the ground truth segmentation mask. For the vessel segmentation experiments, the ground truth centerlines were manually annotated. The evaluation metrics are defined below:

\textit{Dice coefficient} measures the voxelwise overlap between the predicted mask $Y$ and the ground truth mask $G$:
\begin{equation}
\centering
Dice = \frac{2 |Y \cap G|}{|Y| + |G|}
\label{eq:dicecoeff}
\end{equation}

\textit{Centerline completeness} measures the proportion of the length of correctly detected centerlines (i.e., the intersection between the predicted mask $Y$ and the ground truth centerlines $G_{\mathrm{cl}}$) with respect to the length of ground truth centerlines $G_{\mathrm{cl}}$:

\begin{equation}
\centering
Completeness = \frac{|Y \cap G_{\mathrm{cl}}|}{|G_{\mathrm{cl}}|}
\label{eq:completeness}
\end{equation}

\textit{Centerline leakage} measures the proportion of the length of false positive centerlines (i.e., the intersection between the predicted centerlines $Y_{\mathrm{cl}}$ and the ground truth background $1-G$) with respect to the length of ground truth centerlines $G_{\mathrm{cl}}$:

\begin{equation}
\centering
Leakage = \frac{|Y_{\mathrm{cl}} \cap (1-G)|}{|G_{\mathrm{cl}}|}
\label{eq:leakage}
\end{equation}

\textit{Gaps} measures the number of continuity gaps in the correctly detected centerlines (i.e., the intersection between the predicted mask $Y$ and the ground truth centerlines $G_{\mathrm{cl}}$). It is calculated with connected component analysis~\cite{conncompon1996} as follows:
\begin{equation}
\centering
Gaps = NCC(Y \cap G_{\mathrm{cl}}) - NCC(G_{\mathrm{cl}})
\label{eq:gaps}
\end{equation}
with $NCC$ counting the number of 26-neighbor-connected components in the input centerlines.

\section{Results}
\label{sec:results}

% Fully-supervised settings
\subsection{Segmentation Results}
\label{sec:fullresults}

\begin{table*}[h]
  \setlength{\tabcolsep}{4pt}
  \caption{\textbf{Results for airway segmentation.} Average performance (standard deviation) over the results obtained from three random data splits. LR: simple label refinement network. LR+Syn(init): label refinement method with synthetic errors on initial segmentations. LR+Syn: label refinement method with synthetic errors on ground truth segmentations. LR+Syn+LASN: label refinement method with label appearance simulation network. \textcolor{black}{$\uparrow$}: significantly better than the U-Net baseline ($p<0.05$). \textcolor{black}{$\downarrow$}: significantly worse than the U-Net baseline ($p<0.05$). \textcolor{black}{$\Uparrow$}: significantly better than the proposed method ($p<0.05$). P-values are calculated by the paired two-sided Student's T-test (on the average results from the three data splits). Boldface: best and not significantly different from the best results (semi-supervised results are not considered).}
  \label{Table:airwayfull}
  \centering
 \begin{tabular}{lllll}
    \toprule
    Methods & \mc{1}{b}{Dice} & \mc{1}{b}{Completeness} & \mc{1}{b}{Leakage} & \mc{1}{b}{Gaps} \\
    \midrule    
    U-Net baseline~\cite{garciauceda2021} & 0.76 (0.05) & 0.74 (0.12) & 0.23 (0.19) & 95.73 (47.94) \\
    %Restart U-Net & 0.7623 (0.05) & 0.7177 (0.12)\textcolor{black}{$\downarrow$} & \textbf{0.1844} (0.18)\textcolor{black}{$\uparrow$} \\
    DoubleU-Net~\cite{doubleunet} & 0.77 (0.05)\textcolor{black}{$\uparrow$} & 0.73 (0.11) & 0.21 (0.18) & 99.93 (48.11) \\
    SCAN~\cite{adv} & 0.77 (0.05)\textcolor{black}{$\uparrow$} & 0.75 (0.11)\textcolor{black}{$\uparrow$} & 0.31 (0.23)\textcolor{black}{$\downarrow$} & 98.83 (48.81) \\
    \midrule    
    LR & 0.76 (0.05) & 0.74 (0.11)\textcolor{black}{$\uparrow$} & 0.23 (0.17) & 94.90 (47.66) \\
    LR+Syn(init) & 0.77 (0.06) & 0.73 (0.12) & 0.19 (0.17) & 94.92 (50.14) \\
    LR+Syn & \textbf{0.79} (0.05)\textcolor{black}{$\uparrow$} & 0.73 (0.12) & \textbf{0.17} (0.17)\textcolor{black}{$\uparrow$} & \textbf{93.54} (50.83)\textcolor{black}{$\uparrow$} \\
    LR+Syn+LASN (proposed) & \textbf{0.79} (0.05)\textcolor{black}{$\uparrow$} & \textbf{0.75} (0.11)\textcolor{black}{$\uparrow$} & 0.20 (0.16)\textcolor{black}{$\uparrow$} & \textbf{91.63} (48.63)\textcolor{black}{$\uparrow$} \\
    \midrule    
    LR+Syn+LASN+Unlabeled & 0.81 (0.04)\textcolor{black}{$\Uparrow$} & 0.77 (0.10)\textcolor{black}{$\Uparrow$} & 0.19 (0.16)\textcolor{black}{$\uparrow$} & 90.53 (48.80)\textcolor{black}{$\uparrow$} \\

    \bottomrule
  \end{tabular}
\end{table*}

The results of our experiments for airway and brain vessel segmentation are shown in Tables~\ref{Table:airwayfull} and~\ref{Table:vesselfull}, respectively. In both applications, the proposed label refinement method achieves the highest Dice and completeness scores, the lowest number of gaps, with a moderate leakage compared to the other methods. This indicates that our method succeeds in learning from the errors in the synthetic labels to correct errors in the real data. In both applications, the baselines with the highest completeness (SCAN for airways and DoubleU-Net for vessels) show a much higher leakage than our method. This indicates that these methods may lack the ability to learn relevant label structural information, and over-segment branches to increase the completeness rather than correcting errors in continuity.

In the ablation study, the label refinement method with synthetic errors (LR+Syn) achieves better Dice, leakage, and number of gaps scores than the baseline refinement network (LR), for both applications. For airway segmentation, the (LR+Syn) method has slightly lower completeness, while this is similar for vessel segmentation. Moreover, adding synthetic errors to the initial segmentations (LR+Syn(init)), in contrast to doing so to the ground truth segmentations (LR+Syn), achieves similar results in all metrics when compared to the baseline U-Net, for both applications. This suggests that the initial segmentations are too incomplete to add sufficient useful synthetic errors to train the refinement network. The proposed method, combining the synthetic errors and the label appearance simulation network (LR+Syn+LASN), achieves a much higher completeness, with similar Dice, leakage and number of gaps scores when compared to the method with only synthetic errors (LR+Syn), for both applications.

\begin{table*}[h]
  \setlength{\tabcolsep}{4pt}
  \caption{\textbf{Results for brain vessel segmentation.} Average performance (standard deviation) over the results obtained from three random data splits. LR: simple label refinement network. LR+Syn(init): label refinement method with synthetic errors on initial segmentations. LR+Syn: label refinement method with synthetic errors on ground truth segmentations. LR+Syn+LASN: label refinement method with label appearance simulation network. \textcolor{black}{$\uparrow$}: significantly better than the U-Net baseline ($p<0.05$). \textcolor{black}{$\downarrow$}: significantly worse than the U-Net baseline ($p<0.05$). \textcolor{black}{$\Uparrow$}: significantly better than the proposed method ($p<0.05$). P-values are calculated by the paired two-sided Student's T-test (on the average results from the three data splits). Boldface: best and not significantly different from the best results (semi-supervised results are not considered).}
  \label{Table:vesselfull}
  \centering
 \begin{tabular}{lllll}
    \toprule
    Methods & \mc{1}{b}{Dice} & \mc{1}{b}{Completeness} & \mc{1}{b}{Leakage} & \mc{1}{b}{Gaps} \\
    \midrule    
    U-Net baseline~\cite{jsu} & 0.57 (0.10) & 0.70 (0.18) & 0.19 (0.18) & 106.68 (161.41) \\
    %Restart U-Net & 0.5595 (0.10)\textcolor{black}{$\downarrow$} & 0.6918 (0.18) & 0.1955 (0.23) \\
    DoubleU-Net~\cite{doubleunet} & 0.59 (0.09)\textcolor{black}{$\uparrow$} & \textbf{0.73} (0.18)\textcolor{black}{$\uparrow$} & 0.18 (0.16) & 92.41 (151.27)\textcolor{black}{$\uparrow$} \\
    SCAN~\cite{adv} & 0.57 (0.09) & 0.70 (0.18) & 0.17 (0.15) & 104.05 (160.91) \\
    \midrule    
    LR & 0.57 (0.10) & 0.70 (0.18) & 0.16 (0.16)\textcolor{black}{$\uparrow$} & 82.05 (139.45)\textcolor{black}{$\uparrow$} \\
    LR+Syn(init) & 0.58 (0.11) & 0.71 (0.19) & 0.18 (0.15) & 69.91 (126.89)\textcolor{black}{$\uparrow$} \\
    LR+Syn & 0.60 (0.11)\textcolor{black}{$\uparrow$} & 0.71 (0.19) & \textbf{0.12} (0.11)\textcolor{black}{$\uparrow$} & 64.86 (115.21)\textcolor{black}{$\uparrow$} \\
    LR+Syn+LASN (proposed) & \textbf{0.62} (0.10)\textcolor{black}{$\uparrow$} & \textbf{0.74} (0.20)\textcolor{black}{$\uparrow$} & 0.14 (0.11)\textcolor{black}{$\uparrow$} & \textbf{46.64} (76.57)\textcolor{black}{$\uparrow$} \\
    \midrule    
    LR+Syn+LASN+Unlabeled & 0.63 (0.09)\textcolor{black}{$\Uparrow$} & 0.75 (0.18)\textcolor{black}{$\uparrow$}& 0.13 (0.11)\textcolor{black}{$\uparrow$} & 42.45 (71.26)\textcolor{black}{$\Uparrow$} \\

    \bottomrule
  \end{tabular}
\end{table*}

\subsection{Semi-supervised Results}
\label{sec:semiresults}

We conducted experiments using semi-supervised learning to train the proposed label refinement method, to investigate the benefit of using additional unlabeled data for training. As labels in which to synthesize errors for the unlabeled data, we used segmentation results on the same data obtained by the proposed method (LR+Syn+LASN) trained on the labeled data. We denote these results as ``pseudo labels''. The error generation in these pseudo labels follows the same strategy and hyperparameters as in the previous experiments (Sections~\ref{sec:errors} and~\ref{sec:parameters}). The pseudo labels are also used as ground truth segmentations for the unlabeled images. These unlabeled data together with the labeled data in the previous experiments are then used to train a new label refinement network.

The results of our semi-supervised experiments for airway and brain vessel segmentation are shown in the last rows in Tables~\ref{Table:airwayfull} and~\ref{Table:vesselfull}, respectively. Adding unlabeled data for training significantly improves the Dice score while the leakage remains similar, for both applications. For airway segmentation, the completeness is also improved, while this is similar for vessel segmentation. This suggests that for vessels, the labeled data provides enough information for the method to obtain segmentations with good completeness. For vessels, the number of gaps is also improved.

\subsection{Influence of the Synthetic Error Rate}
\label{sec:influence}

\begin{figure}[h]
    \centering
    \includegraphics[width=\linewidth]{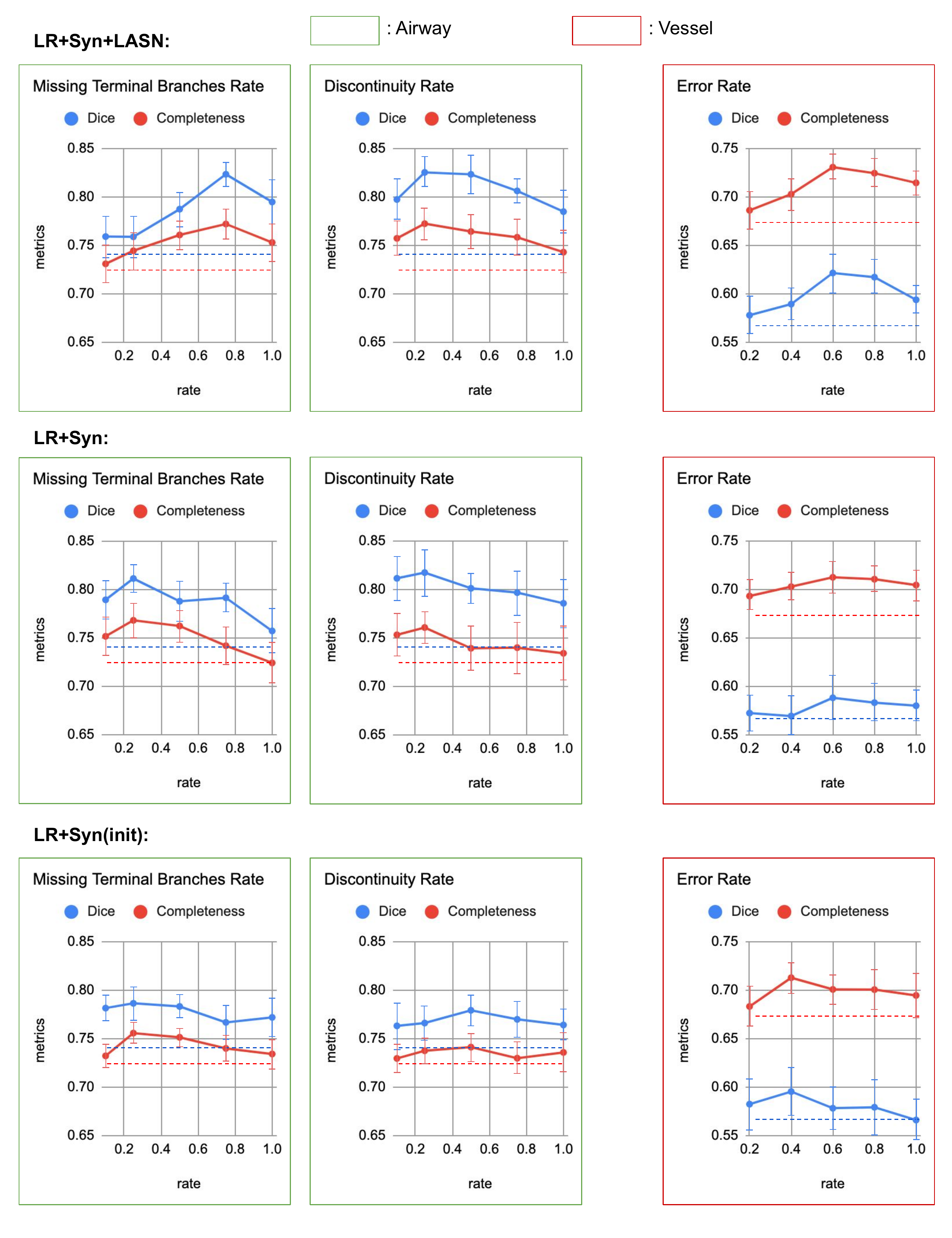} % TODO convert to vector images?
    \caption{\textbf{Influence of the hyperparameters of the proposed method, the maximum synthetic error rates, in the method performance, for airway and brain vessel segmentation.} Results are shown as average performance with standard deviation (error bars), for Dice and completeness metrics, over three random data splits. The results for the baseline (LR) are displayed as dashed line. Better viewed in color.}
    \label{fig:influence}
\end{figure}

The results of our experiments varying the maximum synthetic error rates (Section~\ref{sec:parameters}) are shown in Figure~\ref{fig:influence}. For airway segmentation, with a smaller amount of ``discontinuity'' errors (0.1) the completeness is increased. Between 0.1 and 0.5, changing the amount of ``discontinuity'' errors in the synthetic labels does not affect much the method performance. In contrast, increasing the amount of ``missing terminal branches'' errors improves both Dice and completeness scores, reaching a peak when the maximum error rate is $\approx$ 0.75. This supports our hypothesis that missing terminal branches are relevant errors to be corrected in the initial airway segmentations. For vessel segmentation, a moderate amount (0.6) of ``discontinuity'' errors has a positive effect in the method performance.  
 
When compared to the LR+Syn and LR+Syn(init) methods, the proposed label refinement method is able to learn from higher amounts of synthetic errors, thereby improving the label refinement performance.

\section{Discussion}

In this paper, we propose a novel label refinement method that can correct errors in the initial segmentations from a standard deep segmentation network such as the U-Net. The novelty of our method is that it uses labels augmented with realistic, synthetic errors as training samples, from where the label refinement network can learn to correct the errors. The synthetic errors are automatically generated to simulate common errors observed in the initial segmentations, and are then refined by a label appearance simulation network to resemble the appearance of real errors in the initial segmentations.

We evaluated our method on the segmentation of airways from chest CT scans and brain vessels from CTA images of the brain. In both applications, our method achieved significantly higher Dice overlap and completeness scores, with lower number of gaps and a comparable leakage, when compared to the baseline U-Net and other previous label refinement methods. When segmenting branching structures, a higher completeness means that more and/or longer branches are detected, especially the smaller ones which are challenging to segment.

The ability of our method to segment highly complete tree-like structures with more branches is clinically important, as this could lead to more sensitive biomarkers. For example, in airway analysis, the airway-artery ratio~\cite{kuo2017} and airway tapering~\cite{kuo2020} measures can be used to assess cystic fibrosis lung disease, and including more measurements from the smaller peripheral branches can allow earlier detection of the disease~\cite{tiddens2010}. Moreover, the ability of our method to correct errors in continuity and thereby connect the segmentation is beneficial, as most methods to measure branches assume a fully connected segmentation and discard branches after a discontinuity.

The proposed method outperformed the state-of-the-art label refinement methods DoubleU-Net~\cite{doubleunet} and SCAN~\cite{adv}. Moreover, using semi-supervised learning techniques to train our method with additional unlabeled data we can further improve the method performance, when compared to the fully supervised setting.

\subsection{Comparison to Other Label Refinement Methods}

The main difference between the proposed method and other label refinement methods is the use of a training dataset that includes labels augmented with synthetic errors. Instead of synthetic errors, the DoubleU-Net~\cite{doubleunet} method uses the original images masked by the initial segmentations to train the second network. Although the increased model capacity of DoubleU-Net may improve the segmentations, its ability to correct the errors may be limited by the fact that no new errors are introduced to the input of the second network. This makes it less efficient to implicitly exploit the label structural information similar to a standard U-Net. The SCAN~\cite{adv} method refines the segmentation by making it indistinguishable from the ground truth segmentation through an adversarial loss, where the distribution of the learned features may also provide the general structural information of the objects to be segmented. SCAN mainly focuses on simulating the appearance of the ground truth segmentations. However, SCAN is not designed to learn to correct structural errors explicitly, thus it may not capture the local continuity information as efficiently as our method. This is reflected by the significantly worse completeness reported by SCAN in Tables~\ref{Table:airwayfull} and~\ref{Table:vesselfull}, for both applications. Our method provides an implicit way to enhance the network awareness of the structural information in the ground truth segmentations. For example, after seeing many errors in continuity, the refinement network is expected to understand the local continuity within elongated structures, and consequently to be able to correct these errors in the initial segmentations.

\subsection{Synthetic Errors for Semi-supervised Learning}

With the proposed method, synthetic errors can be added to any pseudo labels obtained on unlabeled data, to be used in semi-supervised learning. In Section~\ref{sec:semiresults} we have shown that our method performance was significantly improved when using additional unlabeled data for training. Our approach to generate synthetic errors could be used together with other common semi-supervised approaches using pseudo labels, e.g., to optimize the prediction consistency of the same image from different models~\cite{tarvainen2017mean}, or the prediction consistency of the same image with different transformations~\cite{bortsova2019semi}. Using synthetic errors in these methods may improve the segmentation quality of pseudo labels from the unlabeled data, which could provide more informative features from these data and thereby improve the method performance.

\subsection{Importance of Realistic Synthetic Errors}

The proposed label refinement network may underperform if the synthetic labels with errors used for training are too different from the initial segmentations. In our method, the synthetic errors are added to the ground truth segmentations, which have a fine and smooth appearance. In contrast, the initial segmentations are more irregular. Our proposed solution is to use a label appearance simulation network trained with an adversarial loss in order to make the appearance of the synthetic labels resemble that of the real initial segmentations. The results in Tables~\ref{Table:airwayfull} and~\ref{Table:vesselfull} clearly show the benefit of using the LASN network in our method. In both applications, without the LASN network could our method (LR+Syn) only slightly improve the segmentation performance with a reduced leakage, when compared to the baseline (LR). This may be due to the positive regularization effect of increasing the variety in the training data by including the synthetic labels with errors. Only after introducing the LASN network was our method able to improve the completeness while retaining an adequate leakage.

\subsection{Applications to Other Segmentation Tasks}

The proposed label refinement method via error synthesis can be applied to other segmentation tasks. The core step is to identify common types of errors in the initial segmentations. For example, a common error we observed in prior work using the U-Net for the segmentation of the aorta and pulmonary arteries from chest CT scans~\cite{chen2021end} was that the segmentation of one of the structures often leaked into the other one, while being both independent anatomical structures. This is mostly due to the obscured boundaries of both arteries on the CT scan. This type of error can be simulated by locally removing the boundaries between the aorta and pulmonary artery classes. Applying our method to correct such errors may improve the overall segmentation performance for this application.

\subsection{Limitations of The Proposed Method}

The main limitation of the proposed method lies in the two-step design and implementation: 1) analyze the errors in the initial segmentations to identify the relevant types of errors, and 2) design and generate the synthetic errors based on these results. The first step requires observation and interpretation by the developers. The synthetic errors we used in this paper are suitable for the segmentation of tree-like structures. However, the relevant types of errors generally differ across different applications and datasets, and therefore the synthetic errors we used are not directly applicable to other segmentation tasks. The second step is typically a complex image processing task. Nevertheless, once the synthetic errors are successfully designed for a given application, the training of our label refinement method can be done fully automatically. 

A limitation of our validation of the proposed method is that we considered only two types of false negative errors (i.e., missing terminal branches and errors in continuity). We did not consider false positive errors because these were much less frequent in the initial segmentations and often appeared as disconnected blobs that can be easily removed without the need for more complex label refinement. Nevertheless, from the results obtained in this paper we expect that our method can successfully correct other types of errors as well.

\section{Conclusion}
We presented a novel label refinement method that is able to learn from synthetic errors to refine the initial segmentations from a base segmentation network. A label appearance simulation network was applied to reduce the appearance difference between the synthetic labels and the real initial segmentations, thereby improving the generalizability of our method. On two segmentation tasks for branching structures, the proposed method achieved a significantly higher Dice overlap and centerline completeness, together with an improved continuity, when compared to previous label refinement methods. The segmentation performance of our method was further improved by using additional unlabeled data for training with semi-supervised learning techniques.

\section*{Acknowledgments}

%This work was partially funded by Chinese Scholarship Council (File No.201706170040), Netherlands Organisation for Scientific Research (NWO) project VI.C.182.042, the MARBLE project (EFRO/OP-Oost: PROJ-00887), the Contrast project (Dutch Heart Foundation (CVON2015-01: CONTRAST), the Brain Foundation Netherlands (HA2015.01.06) and additional funding by the Ministry of Economic Affairs by means of the PPP Allowance made available by the Top Sector Life Sciences \& Health to stimulate public-private partnerships (LSHM17016)).
We want to thank Thirona and H. Tiddens, M. van de Corput and M. Bonte (Erasmus MC - Lung Analysis group) for providing the airway segmentations and anonymous chest CT data used in this study.

%\bibliographystyle{IEEEtran}
%\bibliography{biblio}

\begin{thebibliography}{10}
\providecommand{\url}[1]{#1}
\csname url@samestyle\endcsname
\providecommand{\newblock}{\relax}
\providecommand{\bibinfo}[2]{#2}
\providecommand{\BIBentrySTDinterwordspacing}{\spaceskip=0pt\relax}
\providecommand{\BIBentryALTinterwordstretchfactor}{4}
\providecommand{\BIBentryALTinterwordspacing}{\spaceskip=\fontdimen2\font plus
\BIBentryALTinterwordstretchfactor\fontdimen3\font minus
  \fontdimen4\font\relax}
\providecommand{\BIBforeignlanguage}[2]{{%
\expandafter\ifx\csname l@#1\endcsname\relax
\typeout{** WARNING: IEEEtran.bst: No hyphenation pattern has been}%
\typeout{** loaded for the language `#1'. Using the pattern for}%
\typeout{** the default language instead.}%
\else
\language=\csname l@#1\endcsname
\fi
#2}}
\providecommand{\BIBdecl}{\relax}
\BIBdecl

\bibitem{long2015fully}
J.~Long, E.~Shelhamer, and T.~Darrell, ``Fully convolutional networks for
  semantic segmentation,'' in \emph{Proceedings of the IEEE conference on
  computer vision and pattern recognition}, 2015, pp. 3431--3440.

\bibitem{huang2017densely}
G.~Huang, Z.~Liu, L.~Van Der~Maaten, and K.~Q. Weinberger, ``Densely connected
  convolutional networks,'' in \emph{Proceedings of the IEEE conference on
  computer vision and pattern recognition}, 2017, pp. 4700--4708.

\bibitem{ronneberger2015u}
O.~Ronneberger, P.~Fischer, and T.~Brox, ``U-net: Convolutional networks for
  biomedical image segmentation,'' in \emph{International Conference on Medical
  image computing and computer-assisted intervention}.\hskip 1em plus 0.5em
  minus 0.4em\relax Springer, 2015, pp. 234--241.

\bibitem{UNetreview}
N.~Siddique, S.~Paheding, C.~P. Elkin, and V.~Devabhaktuni, ``U-net and its
  variants for medical image segmentation: A review of theory and
  applications,'' \emph{IEEE Access}, vol.~9, pp. 82\,031--82\,057, 2021.

\bibitem{isensee2021nnu}
F.~Isensee, P.~F. Jaeger, S.~A. Kohl, J.~Petersen, and K.~H. Maier-Hein,
  ``nnu-net: a self-configuring method for deep learning-based biomedical image
  segmentation,'' \emph{Nature methods}, vol.~18, no.~2, pp. 203--211, 2021.

\bibitem{garciauceda2021}
A.~Garcia-Uceda, R.~Selvan, Z.~Saghir, H.~Tiddens, and M.~de~Bruijne,
  ``Automatic airway segmentation from computed tomography using robust and
  efficient 3-d convolutional neural networks,'' \emph{Scientific Reports},
  vol.~11, no.~1, p. 16001, 2021.

\bibitem{jsu}
J.~Su, L.~Wolff, A.~C.~M. van Es, W.~van Zwam, C.~Majoie, D.~W. Dippel,
  A.~van~der Lugt, W.~J. Niessen, and T.~Van~Walsum, ``Automatic collateral
  scoring from 3d cta images,'' \emph{IEEE transactions on medical imaging},
  vol.~39, no.~6, pp. 2190--2200, 2020.

\bibitem{doubleunet}
D.~Jha, M.~A. Riegler, D.~Johansen, P.~Halvorsen, and H.~D. Johansen,
  ``Doubleu-net: A deep convolutional neural network for medical image
  segmentation,'' in \emph{2020 IEEE 33rd International Symposium on
  Computer-Based Medical Systems (CBMS)}.\hskip 1em plus 0.5em minus
  0.4em\relax IEEE, 2020, pp. 558--564.

\bibitem{adv}
W.~Dai, N.~Dong, Z.~Wang, X.~Liang, H.~Zhang, and E.~P. Xing, ``Scan: Structure
  correcting adversarial network for organ segmentation in chest x-rays,'' in
  \emph{Deep learning in medical image analysis and multimodal learning for
  clinical decision support}.\hskip 1em plus 0.5em minus 0.4em\relax Springer,
  2018, pp. 263--273.

\bibitem{labelrefine}
Y.~Yang, Z.~Wang, J.~Liu, K.-T. Cheng, and X.~Yang, ``Label refinement with an
  iterative generative adversarial network for boosting retinal vessel
  segmentation,'' \emph{arXiv preprint arXiv:1912.02589}, 2019.

\bibitem{araujo}
R.~J. Ara{\'u}jo, J.~S. Cardoso, and H.~P. Oliveira, ``A deep learning design
  for improving topology coherence in blood vessel segmentation,'' in
  \emph{International Conference on Medical Image Computing and
  Computer-Assisted Intervention}.\hskip 1em plus 0.5em minus 0.4em\relax
  Springer, 2019, pp. 93--101.

\bibitem{graham2010}
M.~Graham, J.~Gibbs, D.~Cornish, and W.~Higgins, ``Robust {3-D} airway tree
  segmentation for image-guided peripheral bronchoscopy,'' \emph{IEEE
  Transactions on Medical Imaging}, vol.~29, no.~4, pp. 982--997, 2010.

\bibitem{lo2010}
P.~Lo, J.~Sporring, H.~Ashraf, J.~Pedersen, and M.~de~Bruijne, ``Vessel-guided
  airway tree segmentation: A voxel classification approach,'' \emph{Medical
  image analysis}, vol.~14, no.~4, pp. 527--538, 2010.

\bibitem{lo2009}
P.~Lo, J.~Sporring, J.~Pedersen, and M.~de~Bruijne, ``Airway tree extraction
  with locally optimal paths,'' \emph{Medical Image Computing and
  Computer-Assisted Intervention MICCAI}, pp. 51--58, 2009.

\bibitem{exact2012}
P.~Lo, B.~van Ginneken, J.~Reinhardt, Y.~Tarunashree, P.~de~Jong, B.~Irving,
  C.~Fetita, M.~Ortner, R.~Pinho, J.~Sijbers, M.~Feuerstein, A.~Fabijanska,
  C.~Bauer, R.~Beichel, C.~Mendoza, R.~Wiemker, J.~Lee, A.~Reeves, S.~Born,
  O.~Wein-heimer, E.~van Rikxoort, J.~Tschirren, K.~Mori, B.~Odry, D.~Naidich,
  I.~Hart-mann, E.~Hoffman, M.~Prokop, J.~Pedersen, and M.~de~Bruijne,
  ``Extraction of airways from {CT} ({EXACT'09}),'' \emph{IEEE Transactions on
  Medical Imaging}, vol.~31, no.~11, p. 2093–2107, 2012.

\bibitem{qin2021}
Y.~Qin, H.~Zheng, Y.~Gu, X.~Huang, J.~Yang, L.~Wang, F.~Yao, Y.~Zhu, and
  G.~Yang, ``Learning tubule-sensitive {CNNs} for pulmonary airway and
  artery-vein segmentation in {CT},'' \emph{IEEE Transactions on Medical
  Imaging}, vol.~40, no.~6, pp. 1603--1617, 2021.

\bibitem{cheng2021}
G.~Cheng, X.~Wu, W.~Xiang, C.~Guo, H.~Ji, and L.~He, ``Segmentation of the
  airway tree from chest {CT} using tiny atrous convolutional network,''
  \emph{IEEE Access}, vol.~9, pp. 33\,583--33\,594, 2021.

\bibitem{zheng2021}
H.~Zheng, Y.~Qin, Y.~Gu, F.~Xie, J.~Sun, J.~Yang, and G.~Yang, ``Refined
  local-imbalance-based weight for airway segmentation in {CT},'' in
  \emph{Medical Image Computing and Computer Assisted Intervention - MICCAI
  2021}, 2021, pp. 410--419.

\bibitem{sanches}
P.~Sanchesa, C.~Meyer, V.~Vigon, and B.~Naegel, ``Cerebrovascular network
  segmentation of mra images with deep learning,'' in \emph{2019 IEEE 16th
  International Symposium on Biomedical Imaging (ISBI 2019)}.\hskip 1em plus
  0.5em minus 0.4em\relax IEEE, 2019, pp. 768--771.

\bibitem{livne}
M.~Livne, J.~Rieger, O.~U. Aydin, A.~A. Taha, E.~M. Akay, T.~Kossen,
  J.~Sobesky, J.~D. Kelleher, K.~Hildebrand, D.~Frey \emph{et~al.}, ``A u-net
  deep learning framework for high performance vessel segmentation in patients
  with cerebrovascular disease,'' \emph{Frontiers in neuroscience}, vol.~13,
  p.~97, 2019.

\bibitem{adam}
A.~Hilbert, V.~I. Madai, E.~M. Akay, O.~U. Aydin, J.~Behland, J.~Sobesky,
  I.~Galinovic, A.~A. Khalil, A.~A. Taha, J.~Wuerfel \emph{et~al.},
  ``Brave-net: fully automated arterial brain vessel segmentation in patients
  with cerebrovascular disease,'' \emph{Frontiers in Artificial Intelligence},
  vol.~3, p.~78, 2020.

\bibitem{meijs}
M.~Meijs, A.~Patel, S.~C. van~de Leemput, M.~Prokop, E.~J. van Dijk, F.-E.
  de~Leeuw, F.~J. Meijer, B.~van Ginneken, and R.~Manniesing, ``Robust
  segmentation of the full cerebral vasculature in 4d ct of suspected stroke
  patients,'' \emph{Scientific reports}, vol.~7, no.~1, pp. 1--12, 2017.

\bibitem{chestCTdata}
N.~Bouma, H.~Janssens, E.~Andrinopoulou, and H.~Tiddens, ``Airway disease on
  chest computed tomography of preschool children with cystic fibrosis is
  associated with school-age bronchiectasis,'' \emph{Pediatric Pulmonology},
  vol.~55, no.~1, pp. 141--148, 2020.

\bibitem{jansen}
I.~G. Jansen, M.~J. Mulder, and R.-J.~B. Goldhoorn, ``Endovascular treatment
  for acute ischaemic stroke in routine clinical practice: prospective,
  observational cohort study (mr clean registry),'' \emph{bmj}, vol. 360, 2018.

\bibitem{rodriguez2014venous}
D.~Rodriguez-Luna, D.~Dowlatshahi, R.~I. Aviv, C.~A. Molina, Y.~Silva,
  I.~Dzialowski, C.~Lum, A.~Czlonkowska, J.-M. Boulanger, C.~S. Kase
  \emph{et~al.}, ``Venous phase of computed tomography angiography increases
  spot sign detection, but intracerebral hemorrhage expansion is greater in
  spot signs detected in arterial phase,'' \emph{Stroke}, vol.~45, no.~3, pp.
  734--739, 2014.

\bibitem{roman}
R.~Peter, B.~J. Emmer, A.~C. van Es, and T.~van Walsum, ``Cortical and vascular
  probability maps for analysis of human brain in computed tomography images,''
  in \emph{2017 IEEE 14th International Symposium on Biomedical Imaging (ISBI
  2017)}.\hskip 1em plus 0.5em minus 0.4em\relax IEEE, 2017, pp. 1141--1145.

\bibitem{3dunet}
{\"O}.~{\c{C}}i{\c{c}}ek, A.~Abdulkadir, S.~S. Lienkamp, T.~Brox, and
  O.~Ronneberger, ``3d u-net: learning dense volumetric segmentation from
  sparse annotation,'' in \emph{International conference on medical image
  computing and computer-assisted intervention}.\hskip 1em plus 0.5em minus
  0.4em\relax Springer, 2016, pp. 424--432.

\bibitem{pytorch}
A.~Paszke, S.~Gross, F.~Massa, A.~Lerer, J.~Bradbury, G.~Chanan, T.~Killeen,
  Z.~Lin, N.~Gimelshein, L.~Antiga, A.~Desmaison, A.~Kopf, E.~Yang, Z.~DeVito,
  M.~Raison, A.~Tejani, S.~Chilamkurthy, B.~Steiner, L.~Fang, J.~Bai, and
  S.~Chintala, ``{PyTorch}: An imperative style, high-performance deep learning
  library,'' \emph{Advances in Neural Information Processing Systems}, vol.~32,
  2019.

\bibitem{adamoptime}
D.~Kingma and J.~Ba, ``Adam: A method for stochastic optimization,''
  \emph{ArXiv e-prints}, 2017.

\bibitem{skeletonize1994}
T.~Lee, R.~Kashyap, and C.~Chu, ``Building skeleton models via {3-D} medial
  surface axis thinning algorithms,'' \emph{CVGIP: Graphical Models and Image
  Processing}, vol.~56, no.~6, pp. 462--478, 1994.

\bibitem{conncompon1996}
C.~Fiorio and J.~Gustedt, ``Two linear time union-find strategies for image
  processing,'' \emph{Theoretical Computer Science}, vol. 154, no.~2, pp.
  165--181, 1996.

\bibitem{kuo2017}
W.~Kuo, M.~de~Bruijne, J.~Petersen, K.~Nasserinejad, H.~Ozturk, Y.~Chen,
  A.~Perez-Rovira, and H.~Tiddens, ``Diagnosis of bronchiectasis and airway
  wall thickening in children with cystic fibrosis: Objective airway-artery
  quantification,'' \emph{European Radiology}, vol.~27, no.~11, pp. 4680--4689,
  2017.

\bibitem{kuo2020}
W.~Kuo, A.~Perez-Rovira, H.~Tiddens, M.~de~Bruijne, and N.~C.~C. study group,
  ``Airway tapering: an objective image biomarker for bronchiectasis,''
  \emph{European Radiology}, vol.~30, no.~5, pp. 2703--2711, 2020.

\bibitem{tiddens2010}
H.~Tiddens, S.~Donaldson, M.~Rosenfeld, and P.~Pare, ``Cystic fibrosis lung
  disease starts in the small airways: Can we treat it more effectively?''
  \emph{Pediatric Pulmonology}, vol.~45, no.~2, pp. 107--117, 2010.

\bibitem{tarvainen2017mean}
A.~Tarvainen and H.~Valpola, ``Mean teachers are better role models:
  Weight-averaged consistency targets improve semi-supervised deep learning
  results,'' \emph{Advances in neural information processing systems}, vol.~30,
  2017.

\bibitem{bortsova2019semi}
G.~Bortsova, F.~Dubost, L.~Hogeweg, I.~Katramados, and M.~{de Bruijne},
  ``Semi-supervised medical image segmentation via learning consistency under
  transformations,'' in \emph{International Conference on Medical Image
  Computing and Computer-Assisted Intervention}.\hskip 1em plus 0.5em minus
  0.4em\relax Springer, 2019, pp. 810--818.

\bibitem{chen2021end}
S.~Chen, Z.~S. Gamechi, F.~Dubost, G.~van Tulder, and M.~de~Bruijne, ``An
  end-to-end approach to segmentation in medical images with cnn and
  posterior-crf,'' \emph{Medical Image Analysis}, p. 102311, 2021.

\end{thebibliography}

\end{document}